\documentclass{appolb}
\usepackage{graphicx}
\usepackage{amsmath}
\usepackage{glossaries}
\usepackage{hyperref}
\usepackage{tikz}
\usepackage{pgfplots}
\usepackage{geometry}
\usepackage{xcolor}
\usepackage{braket}

\pgfplotsset{compat=1.18}

\newacronym{ckm}{CKM}{Cabibbo-Kobayashi-Maskawa}
\newacronym{schpt}{SChPT}{Staggered Chiral Perturbation Theory}
\newacronym{hisq}{HISQ}{highly-improved staggered quark}
\newacronym{sm}{SM}{Standard Model}
\newacronym{eft}{EFT}{Effective Field Theory}
\newacronym{efts}{EFTs}{Effective Field Theories}

\definecolor{purple}{rgb}{0.5,0.0,0.5}

\begin{document}

\title{A Precision Test of First Row CKM Unitarity from Lattice QCD%
\thanks{Presented at Excited QCD 2026 Workshop}%
}
\author{Ramón Merino$^a$, 
\address{\footnotesize 
$^a$\textit{Departamento de Física Teórica y del Cosmos, Universidad de Granada, E-18071, Granada, Spain}} 
\\[0.7em]
{\textsf{on Behalf of the Fermilab Lattice and MILC Collaborations}}\\[0.7em]
}
\maketitle
\begin{abstract}
High-precision determinations of Cabibbo-Kobayashi-Maskawa (CKM) matrix elements are essential probes of physics Beyond the Standard Model (BSM). Current precision tests show a deficit in the first row unitarity relation.
At the current level of precision, the only relevant CKM matrix elements that contribute to this test are $|V_{ud}|$ and $|V_{us}|$. Without resorting to nuclear inputs, they can be extracted from the combination of the experimental decay width of kaon and pion leptonic decays, along with the theoretical calculation of their decay constants; combined with the decay width of semileptonic kaon decays, with the computation of the corresponding form factor at zero momentum transfer. We review current efforts by the Fermilab Lattice and MILC collaborations towards a correlated analysis of the lattice inputs needed for this test using Highly Improved Staggered Quarks (HISQ) on the $N_f=2+1+1$ MILC configurations along with Staggered Chiral Perturbation Theory (SChPT) as a functional form for the chiral-continuum limit.
\end{abstract}
  
\section{Introduction}

One of the central aims of flavor physics is to perform high precision computations of physical observables that can be used to test the Standard Model (SM). Of particular interest are those observables that are related to one or more elements of the \gls{ckm} matrix.
They govern charged-current weak interactions, remarkably describing a wide range of flavor-changing processes. Searches for signs of Beyond the Standard Model (BSM) physics often rely on the precise determination of \gls{sm} parameters. Thus, any discrepancy between the value of the \gls{ckm} matrix elements obtained by studying different processes or any signs of its deviation from unitarity could serve as a smoking gun for new physics. Moreover, precise determinations of these parameters can impose strong constraints on different extensions of the \gls{sm}.

In the \gls{sm}, the \gls{ckm} matrix is unitary by construction.  
However, current determinations of the elements in the first row result in a $2-3\sigma$ deficit in the unitarity relation, 

\begin{equation}
  \Delta \equiv |V_{ud}|^{2} + |V_{us}|^{2} + |V_{ub}|^{2} - 1 \; \overset{\rm SM}{=}\; 0 \; .
\end{equation}

At the current level of precision, the only relevant parameters in this test are $|V_{ud}|$ and $|V_{us}|$; hence, this deficit is known as the Cabibbo Angle Anomaly (CAA). In this talk, we briefly review the current status of the determination of these elements, as well as the current efforts by the Fermilab Lattice and MILC collaborations to improve the computation of the non-perturbative inputs necessary for the first-row unitarity test. 

\section{Current status}

The most precise way to extract $|V_{ud}|$ is by measuring the decay width of super-allowed beta decays, based on nuclear $0^+\rightarrow0^+$ transitions. Although the experimental measurements are very precise, the total uncertainty is dominated by theoretical corrections dependent on nuclear structure, which are not fully understood yet. For more information on beta decays, see Ref.~\cite{Hayen:2024xjf}. 

Therefore, alternative determinations of $|V_{ud}|$ have been considered. In particular, the second most precise determination comes from neutron decays. These processes do not suffer from nuclear-structure uncertainties, and the total uncertainty is currently dominated by the experimental precision. Although the experimental error makes neutron decays noncompetitive at the moment, the situation can change in the next few years given the expected experimental improvements. 
Another process worth mentioning is pion beta decay. In addition to being free from nuclear structure uncertainties, radiative corrections in this process are well understood, making it the theoretically cleanest channel for determining $|V_{ud}|$. Although its experimental uncertainty is far from being competitive, dedicated experimental efforts will improve the situation. In particular, the \textit{PIONEER} experiment \cite{PIONEER:2025idw} will aim to reduce the uncertainty in the relevant branching ratio by approximately a factor of $6$.

Regarding $|V_{us}|$, the most precise determination comes from semileptonic kaon decays $K\rightarrow\pi\ell\nu_\ell$ ($K_{\ell3}$). The decay width can be related to the desired quantity and a non-perturbative input, the semileptonic form factor at $q^2=0$, -- see Ref.~\cite{FermilabLattice:2018zqv} 

\begin{equation}
  \Gamma(K\rightarrow \pi\ell\nu) \propto |V_{us}f_{+}(0)|^{2}\, .
\end{equation}

An alternative determination of $|V_{us}|$ based on inclusive hadronic tau decays, although not yet competitive in precision, shows a small tension with the one from kaon decays \cite{ExtendedTwistedMass:2024myu}. 
Interestingly, exclusive hadronic tau determinations \cite{HeavyFlavorAveragingGroupHFLAV:2024ctg} lie closer to the kaon decays value, suggesting that the $|V_{us}|$ picture is not yet well understood.
 
 Lastly, leptonic pion and kaon decays ($K_{\ell2}/\pi_{\ell2}$) can be studied to extract the ratio between the CKM elements $|V_{us}|/|V_{ud}|$. The ratio of the decay widths can be related to the ratio between $|V_{us}|/|V_{ud}|$ and the ratio of the decay constants $f_K/f_\pi$

\begin{equation}
  \frac{\Gamma(K\rightarrow\ell\nu\ell)}{\Gamma(\pi\rightarrow\ell\nu\ell)} \propto \frac{|V_{us}|^2}{|V_{ud}|^2}\frac{f_K^2}{f_\pi^2} \, .
\end{equation}

\begin{figure}[htb]
    \centering
    \includegraphics[width=0.9\linewidth]{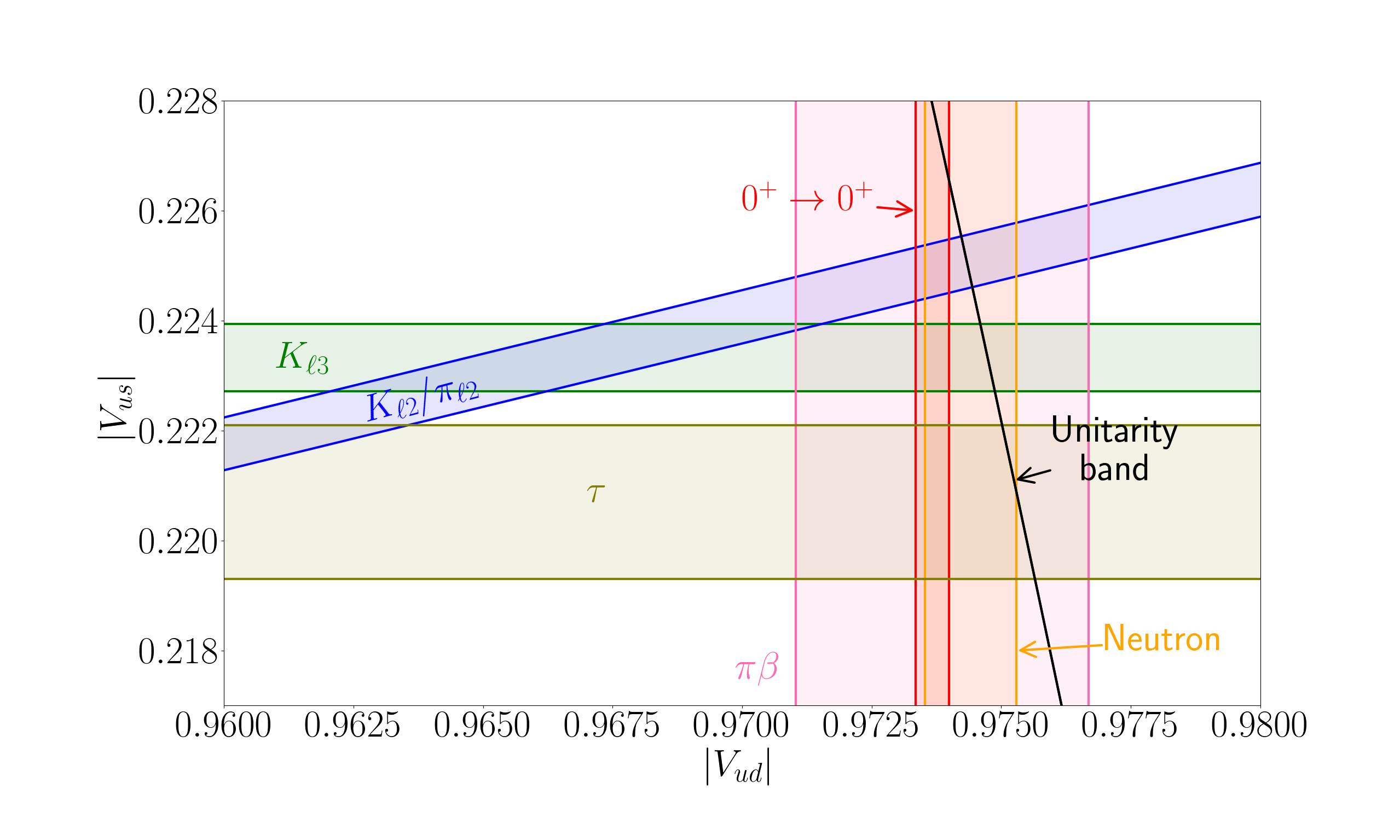}
    \caption{Determinations of $|V_{ud}|$, $|V_{us}|$ and $|V_{us}|/|V_{ud}|$. The black band represents the unitarity constraint, taking into account the value of $|V_{ub}|$ from the PDG~\cite{ParticleDataGroup:2024cfk}. The green band represents the determination of $|V_{us}|$ coming from kaon  decays\cite{FermilabLattice:2018zqv}, while the olive band corresponds to the average of the exclusive and inclusive determinations from tau decays \cite{HeavyFlavorAveragingGroupHFLAV:2024ctg}.
    Red, orange and pink bands show determinations of $|V_{ud}|$ using superallowed beta, neutron and pion beta decays, respectively. They are taken from Ref.~\cite{Cirigliano:2022yyo}. The blue oblique band is the determination of the ratio $|V_{us}|/|V_{ud}|$ from leptonic kaon and pion decays~\cite{Bazavov:2017lyh}. 
    }
    \label{fig:situationVudVus}
\end{figure}

A summary of the different determinations of $|V_{ud}|$ and $|V_{us}|$ is shown in Fig~\ref{fig:situationVudVus}. Remarkably, first-row unitarity is not satisfied for almost any of the combinations of $|V_{ud}|$ and $|V_{us}|$, with deviations at the level of $\sim 1-3\sigma$. Furthermore, there are tensions 
between the different determinations of $|V_{us}|$ coming from different processes.

If we want to perform the first-row unitarity test without relying on nuclear inputs, we can combine the ratio $|V_{us}|/|V_{ud}|$ from kaon and pion leptonic decays, with the value for $|V_{us}|$ from semileptonic kaon decays.
The Fermilab Lattice and MILC collaborations computed the non-perturbative inputs needed for this test: the decay constants ratio $f_K/f_\pi$~\cite{Bazavov:2017lyh} and the form factor $f_+(0)$~\cite{FermilabLattice:2018zqv}, using lattice QCD techniques. We will now discuss our strategy to improve upon those two analyses in order to increase the precision and study correlations among both quantities.

\section{Analysis strategy}\label{sec:new}

Our calculation is carried out using lattice QCD, a first-principles, non-perturbative method based on the numerical evaluation of the path integral. We employ gauge-field configurations generated by the MILC collaboration, covering several lattice spacings, quark masses, and physical volumes. 
These configurations include the effect of the up, down, strange and charm quarks in the sea ($N_f=2+1+1)$. Further details on the ensembles analysed in this work can be found in Refs.~\cite{FermilabLattice:2018zqv,Bazavov:2017lyh,Lattice:2026euh}.  

Physical predictions from lattice QCD require extrapolation to the continuum limit and infinite volume, as well as corrections for unphysical quark masses when simulations are not run at the physical point. In order to do so, 
we use Staggered Chiral Perturbation Theory (SChPT) to build our chiral-continuum fit function. SChPT is a modification of continuum ChPT that enables the systematic inclusion of discretization and chiral effects along with finite volume and strong-isospin-breaking corrections. We perform our chiral-continuum analysis by building a fit function from NLO SChPT plus NNLO ChPT plus additional terms that parameterize higher order discretization and chiral effects.

 In our previous work \cite{Bazavov:2017lyh}, the computation of the ratio $f_K/f_\pi$ relied mainly on the physical quark mass ensembles.
 By switching to an SChPT-based fit function, we are now able to include data at unphysical quark masses in the chiral-continuum analysis. This improves our statistical precision and allows for a more robust analysis in which systematic uncertainties are better controlled.

Since ChPT converges poorly at the strange quark mass, we perform our chiral-continuum analysis of $f_K/f_\pi$ in two separate steps. First, we restrict the analysis to ensembles with a lighter-than-physical strange-quark mass, where SChPT shows good convergence, to obtain a correlated determination of the SChPT Low-Energy Constants (LECs). In the second step, we include all available data, supplementing the fit function with additional higher-order terms, and use the LECs from the first step as priors to perform a precise determination of $f_K/f_\pi$ \cite{Lattice:2026euh}.

\section{Preliminary results}

\begin{figure}[htb]
    \centering
    \includegraphics[width=0.8\linewidth,trim=200 0 0 0,clip]{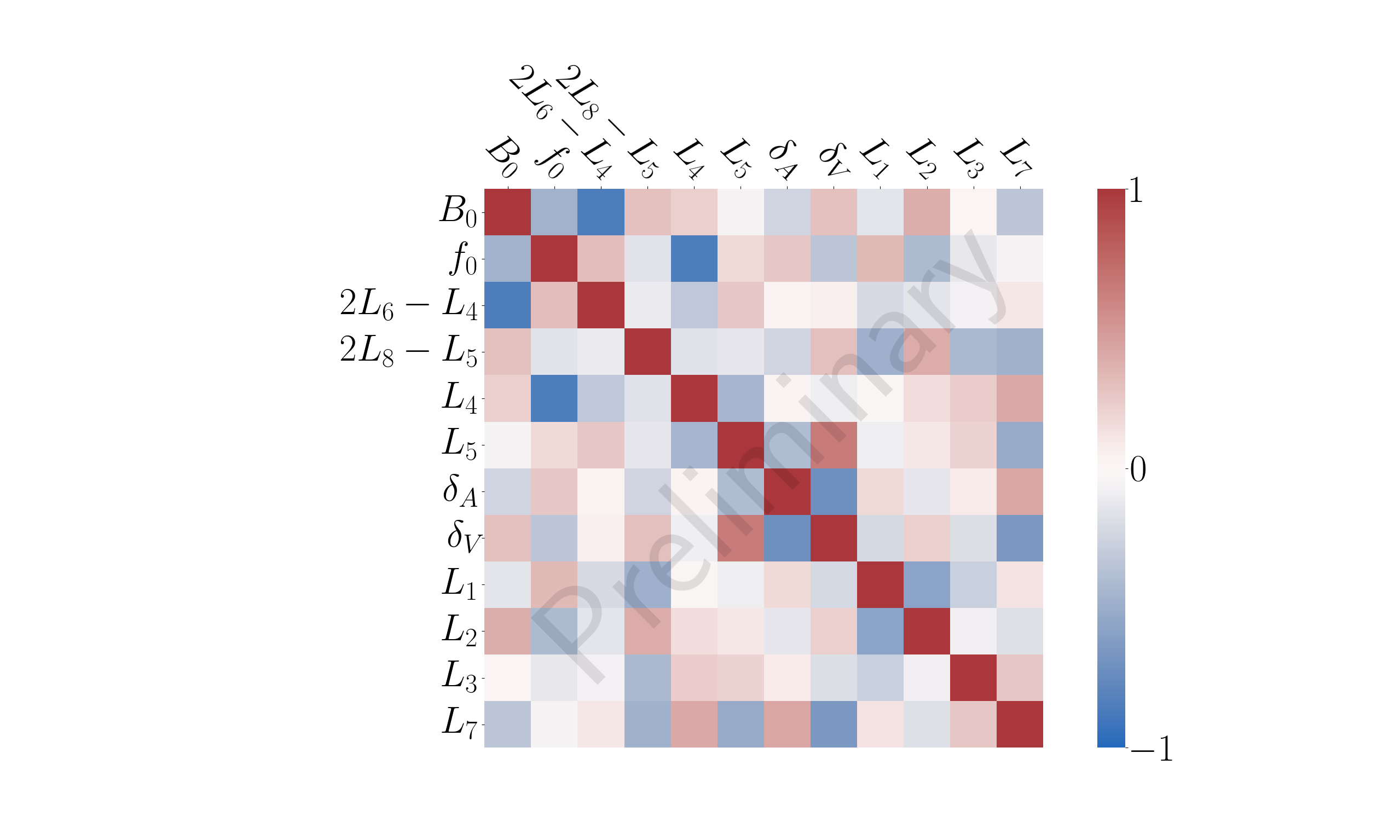}
    \caption{
    Correlation matrix for a representative subset of SChPT LECs obtained from the first step of the chiral-continuum analysis. Several pairs exhibit strong correlations, underlining the importance of a correlated treatment. For more information, see \cite{Lattice:2026euh}.
    }
    \label{fig:corrmatrix}
\end{figure}

We focus our discussion on the chiral-continuum analysis explained in Sec~\ref{sec:new}. To estimate covariances between fit parameters, we perform a resampling analysis in which synthetic samples are drawn from the data covariance matrix.
We perform the first step in the analysis, where only lighter-than-physical data are included, obtaining a correlated distribution of the LECs of SChPT. We obtain precise determinations of certain LEC combinations, with strong correlations between some of them, as shown in Fig.~\ref{fig:corrmatrix}.

In the second step, we use the LEC posteriors from the first step as priors, and then perform the chiral-continuum fit including all data. Since we now include data at heavier masses, it is necessary to include additional terms in the fit function to account for higher order discretization and chiral effects -- see \cite{Lattice:2026euh} for more details. We find the LEC central values and correlations to be consistent in both steps, and we obtain a precise value for $f_K/f_\pi$. However, the inclusion of more parameters in the fit function makes the correlation estimation in the resampling analysis more challenging. We are still working on the best strategy to obtain a robust estimation of correlations between $f_K/f_\pi$ and LECs.

Turning to the semileptonic form factor $f_+(0)$, we can exploit the fact that its SChPT chiral-continuum expression shares common parameters with the one in the decay constants analysis. 
We exploit this by using the LEC distributions from the leptonic analysis as priors in the semileptonic fits, which allows us to propagate correlations between the two quantities.
We find non-zero correlations between the form factor $f_+(0)$ and the LECs, signaling that our resampling strategy is able to carry correlations throughout both analyses -- preliminary results are shown in Ref~\cite{Lattice:2026euh}.

Such a correlated input also has implications for Beyond the Standard Model physics. A recent global SMEFT fit~\cite{Cirigliano:2023nol} showed that first-row CKM tensions can be alleviated by adding operators that modify right-handed charged currents (one possible UV completion that could generate such operators is found in vector-like quark theories). The sensitivity of such analyses to new physics depends directly on the uncertainties and correlations between the SM inputs. We expect that our simultaneous determination of $f_+(0)$ and $f_K/f_\pi$, together with their correlations, could improve such analyses.

\section{Conclusions and outlook}

We have reviewed the current status of first-row CKM unitarity and presented the ongoing efforts by the Fermilab Lattice and MILC collaborations towards the first correlated analysis of $f_K/f_\pi$ and $f_+(0)$. By performing the chiral-continuum analysis, choosing a fit function based on SChPT, our analysis yields not only these two quantities but also a precise and correlated determination of the SChPT LECs.
 Such a correlated determination will provide a more rigorous input to CKM unitarity tests and improve the sensitivity of high-precision searches for new physics.

\section*{Acknowledgments}

Computations for this work were carried out with resources provided by the USQCD Collaboration; by the ALCF and NERSC, which are funded by the U.S. Department of Energy (DOE); and by NCAR, NCSA, NICS, TACC, and Blue Waters, which are funded through the U.S. National Science Foundation (NSF). 

This work was supported in part 
by MICIU/AEI/10.13039/501100011033 under Grant PID2022-140440NB-C21 and 
by the Junta de Andaluc\'{\i}a grant FQM 101.

\bibliography{ref}

\end{document}